\documentclass[9pt,twocolumn,twoside]{opticajnl}
\journal{opticajournal} 

\setboolean{shortarticle}{false}

\usepackage{lineno}
\usepackage{upgreek}
\usepackage{subcaption}

\title{Three-Wave Interaction Grating Coupler with Sub-Decibel Insertion Loss at Normal Incidence}

\author[1,*]{Carson G. Valdez}
\author[2]{Simon A. Bongarz}
\author[1]{Anne R. Kroo}
\author[1]{Anna J. Miller}
\author[2]{Michel J. F. Digonnet}
\author[1]{David A. B. Miller}
\author[1]{Olav Solgaard}

\affil[1]{Electrical Engineering Department, Stanford University, Stanford CA 94305}
\affil[2]{Applied Physics Department, Stanford University, Stanford CA 94305}

\affil[*]{carsongv@stanford.edu}

\begin{abstract}
We report the design, fabrication in a commercial foundry, and experimental results of high-efficiency, normal incidence grating couplers for silicon photonics. We observe a maximum coupling efficiency of 85.4$\%$ (-0.69 dB) with a 1 dB bandwidth of 20 nm at a central wavelength of 1546 nm. These experimental results verify earlier theoretical and simulation results and pave the way for the use of perfectly vertical grating couplers, as an alternative to edge coupling, in silicon photonics applications that are sensitive to input coupling loss. Further, these results enable the use of grating couplers for vertically oriented elements, such as multicore fibers and VCSELs, and address challenges associated with coupling to free space beams.

\end{abstract}

\setboolean{displaycopyright}{false} 

\begin{document}

\maketitle

\section{Introduction}

Silicon photonics is making inroads in optical systems ranging from optical communication \cite{Valdez23,Shekhar2024}, spectroscopy \cite{Li2022,miller2025, Bates2021}, and quantum optics \cite{Psiquantum, Luo2023} to astro-photonics \cite{Sirbu2024, Morgan2025,Glint}. The gold standard for coupling into silicon photonic single mode waveguides from free space or standard single-mode optical fiber is edge coupling, which can achieve both high coupling efficiency and large bandwidths \cite{Marchetti19, Mu2020}. However, edge coupling has several drawbacks. It requires additional preparation of the facets of the silicon photonics waveguides and poses challenges to wafer-level testing, requiring specialized equipment \cite{Yuan2025}. Additionally, in an edge coupling configuration the mode field diameter (MFD) of the integrated waveguide is limited by the thickness of the lower cladding which is typically 2 $\upmu$m to 3 $\upmu$m thick. As a result, efficient mode matching to edge couplers often requires either complex processing to allow for suspended silicon edge couplers \cite{Barwicz2019} or lens tapered fibers, which are expensive, fragile, and impose strict lateral alignment tolerances when compared to larger MFD single mode fibers \cite{Wang2016, Marchetti19, Mu2020}.

Grating couplers provide a solution to many of the challenges associated with edge couplers. Grating couplers may be fabricated anywhere across the surface of a wafer without the need for complex processing. In a grating coupler configuration, the mode is adiabatically tapered within the waveguiding layer such that there are few limitations on the resulting mode field diameter. However, typical single-layer Bragg grating couplers suffer from low coupling efficiency and low bandwidth \cite{Cheng2020, Hooten2020}. Additionally, standard grating couplers are prone to large back reflections when designed to operate at normal incidence, as is desirable for coupling to vertically oriented devices such as VCSELs and multicore fibers \cite{Cheng2020, Hooten2020, Watanabe2017, Tong2018}. This has led to a focus on the development of grating coupler designs that have high coupling efficiency and high bandwidth, which may operate at normal incidence for certain applications. This work builds on earlier analytical and numerical modeling of high efficiency, normal incidence grating couplers by designing, fabricating in a commercial foundry, and experimentally demonstrating grating couplers with a peak coupling efficiency of 85.4$\%$ (-0.69 dB) at normal incidence.

\section{Modeling}
Typical grating couplers enable coupling between integrated waveguides and fibers by periodically perturbing the waveguide mode via etches along the length of the waveguide. Each of these perturbations acts as a scattering site which results in constructive interference of the scattered light at a particular angle $\theta$. The relation between the geometry of the waveguide, grating, and emission/acceptance angle is given by the grating equation \cite{Cheng2020}:

\begin{equation}
\frac{2\pi}{\lambda}n_{gr}-\frac{2\pi m}{\Lambda}=k_0\sin{\left(\theta\right)}
\label{eq:refname1}
\end{equation}

Here $\lambda$ is the operating wavelength, $\Lambda$ is the grating period, $m$ is the diffraction order which we typically take to be 1, $k_0$ is the wavenumber in the cladding, and $n_{gr}$ is the effective index of the grating which is the weighted average of the effective index for the unperturbed and perturbed waveguide modes over the period of the grating. 

The efficiency of such a coupler is then given by the power normalized mode overlap integral of the grating mode and a desired mode \cite{Zhao2020}. In the case where a grating is being treated as an output coupler, with light injected from the integrated waveguide and emitted toward a fiber, the efficiency is expressed as:

\begin{equation}
\eta=\frac{1}{4P_0P_{fib}}\left|\iint_{S}E_{grat}\times H_{fib}^\ast\cdot d S\right|^2
\label{eq:refname2}
\end{equation}

Here $P_0$ is the total power entering and leaving the system, $P_{fib}$ is the power in the simulated ideal fiber mode, $E_{grat}$ is the electric field profile emitted from the grating, and $H_{fib}$ is the magnetic field profile of the fiber mode. For coupling to a single mode fiber at 1550 nm, the desired fiber mode is often taken to be a Gaussian profile with a MFD of 10.4 $\upmu$m. As a passive optical component, the grating maintains reciprocity and will yield the same efficiency when treated as an input coupler.

It is often useful to consider the coupling efficiency as the product of two figures: the directionality and the mode overlap. Directionality is the ratio of power directed in the intended propagation direction to the total injected power. The mode overlap is a measure of the similarity between the emitted grating field and the desired fiber mode. It has been established \cite{Cheng2020,Zhao2020} that a good mode overlap can be achieved in single-layer gratings by apodizing the scattering strength along the length of a grating. This is achieved by varying the width of the etched perturbations systematically. Conversely, the directionality of a single-layer grating faces a limitation. In a single-layer grating where the etch depth matches the thickness of the waveguide layer, vertical symmetry dictates that half of the power be directed downward toward the substrate. Partial etching has been employed to alleviate this limitation. However, on a 220 nm thick SOI platform, this technique has been shown to have an upper limit of 65$\%$ coupling efficiency \cite{Bozzola2015,Marchetti2017}.

Other methods have been introduced to break the vertical symmetry of the system such as bottom side reflectors (both Bragg \cite{Zhang2019} and metallic \cite{Laere2007}), as well as various material overlays \cite{Vermeulen2010,Tong2018} and bilayers \cite{Hooten2022}. Each of these techniques works by introducing at least one new material layer to the system, requiring additional complex processing.

Directionality is particularly challenging for gratings designed to operate at normal incidence. This is a result of strong coupling to higher order (m>1) reflections back into the waveguide \cite{Cheng2020}. Multi-etch step gratings have been shown \cite{Benedikovic2015, Chen2017, Watanabe2017, Dezfouli2020, Tian2022, Zhou2022, Zhou2023, Naoki2025} to improve directionality at normal incidence, by introducing multiple scattering sites within a local period. The additional degrees of freedom introduced by multiple scattering sites enable control over the interference between multiple backward scattered waves from a single local period. The same principle holds true for waves scattered toward the substrate. Designing for the simultaneous destructive interference of these losses ensures good directionality.

\begin{figure}[ht]
\centering
\includegraphics[width=\linewidth]{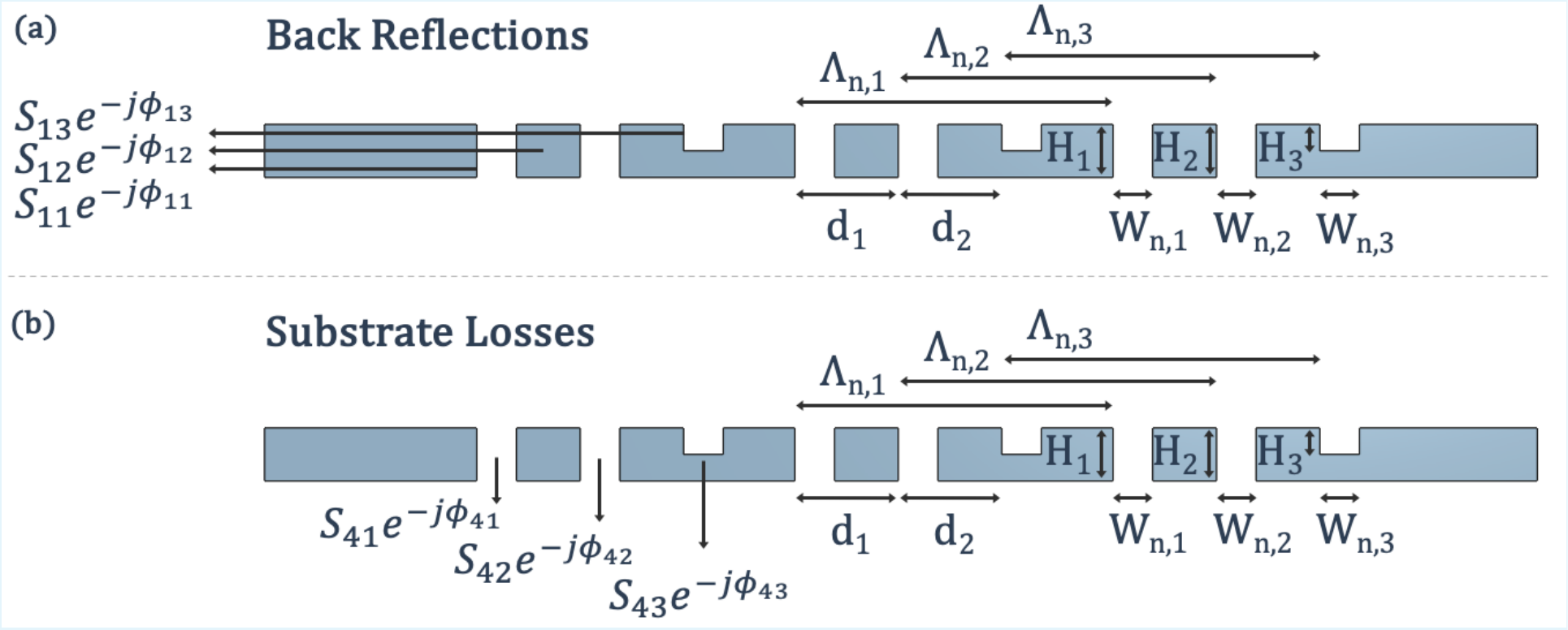}
\caption{Generalized geometry of a three-wave interaction grating parameterized by $W_{n,m}$, $\Lambda_{n,m}$, $H_m$, $d_1$ and $d_2$. $W_{n,m}$ represents the width of the $m^{th}$ scatterer in the $n^{th}$ local period. $\Lambda_{n,m}$ represents the separation between the $m^{th}$ scatterers in the $n$ and $n+1$ local periods. $H_m$ is the etch depth of the $m^{th}$ scattering site across all local periods. $d_1$ and $d_2$ are initial offset values between the scattering sites in the first local period. Panels \textbf{a} and \textbf{b} show the back reflected and downward scattered waves from three scattering sites respectively. }
\label{fig:Geometry}
\end{figure}

Prior work \cite{Michaels2018, Notaros2016, Dai2015, Zhou2022, Zhou2023} has shown that high coupling efficiency at normal incidence can be achieved with as few as two scattering sites per local period; however, it imposes requirements on the geometry of the structure which are challenging to fabricate. Namely, it requires the existence of buried voids beneath the top surface of the silicon waveguide. While this has been demonstrated in practice, it requires the use of an additional polysilicon layer \cite{Notaros2016, Dai2015, Zhou2022, Zhou2023}. It has been further demonstrated that these restrictions to device geometry can be overcome by the inclusion of a third scattering site within each local period, as shown in figure \ref{fig:Geometry}. These three-wave interaction gratings (TWIGs) entirely remove the need for additional material layers without introducing any additional fabrication steps \cite{Valdez2024}.

Here we use the methods developed in \cite{Valdez2024} to determine the design parameters of a TWIG coupler optimized for the material stack provided by the commercial foundry Applied Nanotools. We first determine an initial device geometry by applying the analytical solutions to a subset of the TWIG geometry derived in \cite{Valdez2024}. We then employ adjoint method based inverse design to optimize the geometry using our analytical solution as an initial condition. Here we use the open-source software EMOPT to simulate and optimize the two-dimensional dielectric distribution of the TWIG coupler via FDTD simulations. Two-dimensional FDTD's are sufficient to approximate a three-dimensional grating coupler of sufficient width in the out-of-plane dimension. Within the optimization, we will enforce certain conditions on the figure of merit (FOM) such that we balance the trade-offs between coupling efficiency, bandwidth, and fabricability.

The specific subset of geometries solved for in \cite{Valdez2024} sets $\Lambda_{n,1} = \Lambda_{n,2} =\Lambda_{n,3} =\Lambda$ and $W_{n,1} = W_{n,2} = W_{n,3} = W$ where $W= (1-D)\Lambda$ for all n, defining a uniform TWIG grating. Here D $\in$ [0,1] is the ratio of a local period that is not etched. Under these conditions, the grating equation may be rewritten as: 

\begin{equation}
\Lambda=\frac{m\lambda}{\left(1-D\right)\left(n_1+n_2+n_3\right)+\left(3D-2\right)n_w-n_0sin\left(\theta\right)}
\label{eq:refname3}
\end{equation}

\begin{figure*}[ht]
\centering
\includegraphics[width=\linewidth]{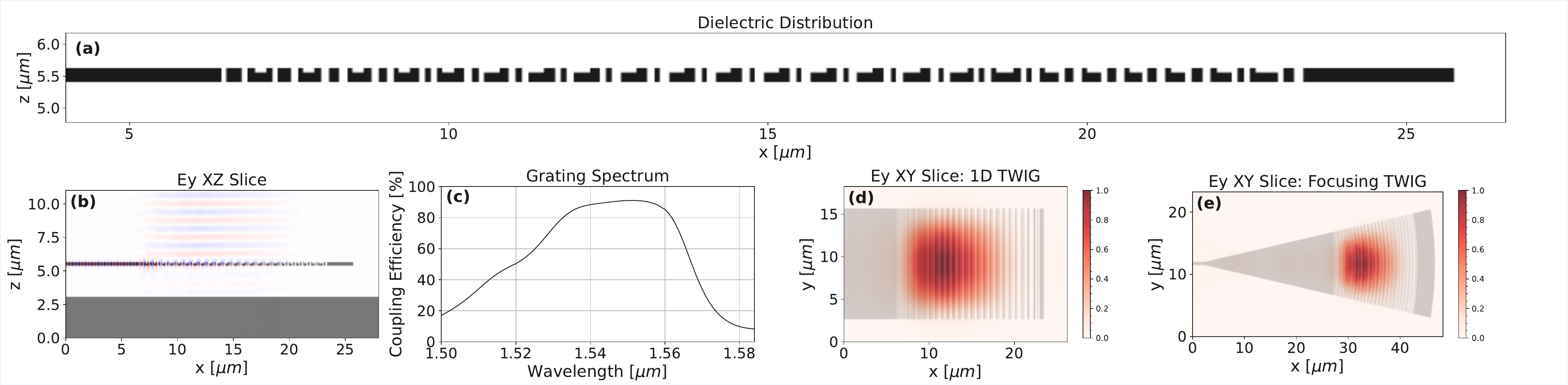}
\caption{Simulation results of TWIG coupler optimization. \textbf{(a)} Optimized geometry demonstrating apodization of the etch widths to produce a gaussian mode profile. The background index here is associated with silicon dioxide (1.444). \textbf{(b)} Electric field profile associated with the quasi-TE mode of the grating. The output is directed perfectly vertical with minimal back reflections and substrate losses. \textbf{(c)} Coupling spectra of the optimized coupler with a peak efficiency of 91 $\%$ and a 1 dB bandwidth of 34 nm spanning the c-band. Field profiles generated by a \textbf{(d)} one-dimensional and \textbf{(e)} focusing TWIG coupler.}
\label{fig:Simulations}
\end{figure*}

Where $n_1$, $n_{2}$, and $n_3$ represent the effective index in the etched regions with etch depths $H_1$, $H_2$, and $H_3$ respectively, $n_w$ represents the effective index of the unperturbed waveguide, and $n_0$ represents the index of the cladding material. By adjusting the periodicity $\Lambda$, the gratings may be designed for arbitrary emission/acceptance angles $\theta$. In order to achieve coupling at normal incidence ($\theta = 0$) we must impose phase matching conditions to ensure destructive interference between the back reflections of each scattering site within a local period. The necessary widths $d_1$ and $d_2$ are found by equations \ref{eq:refname4} and \ref{eq:refname5}  respectively, where $\phi_{ij}$ denotes the relative phase difference between back reflected waves from scattering sites $i$ and $j$ \cite{Valdez2024}.

\begin{equation}
d_1=\frac{\left(\phi_{12}\right)\lambda}{4\pi n_w}+\Lambda\left(1-D\right)\left(1-\frac{n_1}{n_w}\right)
\label{eq:refname4}
\end{equation}

\begin{equation}
d_2=\frac{\left(\left(\phi_{13}-\phi_{12}\right)\right)\lambda}{4\pi n_w}+\Lambda\left(1-D\right)\left(1-\frac{n_2}{n_w}\right)
\label{eq:refname5}
\end{equation}

Together equations \ref{eq:refname3}, \ref{eq:refname4}, and \ref{eq:refname5} completely parameterize the geometry of a TWIG coupler of a given material stack with selected etch depths $H_1$, $H_2$, and $H_3$. The material stack available to us for this work uses a 725 $\upmu$m thick silicon handle wafer with a 2 $\upmu$m thick thermally grown buried oxide layer. The silicon waveguide layer is a standard 220 nm thick device layer. An additional 2.5 $\upmu$m of oxide cladding is deposited above the waveguide layer following patterning. For the etch depths we have selected $H_1$ = $H_2$ = 220 nm. We have set these two depths equal such that they can be realized in the same process step and minimize the number of lithography steps needed. We have decided to fully etch through the waveguiding layer to remove any possibility of an over-etching error. The third scattering site is assigned a partial etch depth of $H_3$ = 70 nm. It must be noted that these etch depths were assigned somewhat arbitrarily and that the TWIG geometry can accommodate any number combinations for etch depths \cite{Valdez2024}.

The parameters used as the initial condition for the inverse design of this TWIG coupler are given in table 1 below. Prior to optimization, we assume that each local period of the grating is identical such that there is no apodization. 

\begin{table}[htbp]
\small
\centering
\caption{\bf Initial Conditions for Optimization}
\begin{tabular}{c c c c c c c}
\hline
$\Lambda$ & $D$  & $H_1$ & $H_2$ & $H_3$ & $d_1$ & $d_2$  \\
\hline
$637$ nm & $85\%$ & $220$ nm & $220$ nm & $70$ nm & $194$ nm & $246$ nm \\
\hline
\end{tabular}
  \label{tab:intialization}
\end{table}

\section{Optimization}

The fully arbitrary TWIG geometry is described by $3N$ local periodicities, $3N$ etch widths, and 3 global height parameters where $N$ is the number of local periods. In the devices designed in this work $N = 24$. Optimization of such devices requires fine-tuning in a parameter space that is nearly 150 dimensions. Adjoint method-based optimization allows us to efficiently navigate the parameter space by calculating the gradient of the FOM relative to each parameter using a single forward and backward simulation \cite{Michaels2018,Michaels2020}. By using the geometry described above as an initialization for our optimization, we can reduce the likelihood of converging on a false local optimum. 

To ensure a reasonable balance between bandwidth and coupling efficiency, the figure of merit is defined as the average coupling efficiency over several wavelengths of interest. We have chosen to discretize the FOM over only a few wavelengths of interest to reduce the computational load of each iteration in the optimization. Prior work has shown that the method is effective at improving bandwidth with small impact to the peak coupling efficiency \cite{Valdez2024}. Additionally, we impose two design restrictions with regards to the minimum feature size of the TWIG coupler based on the tolerances provided by Applied Nanotools. It is enforced that all features with a full etch depth of 220 nm have a minimum critical dimension (etch width) of 70 nm and that all features with a partial etch depth of 70 nm have a minimum critical dimension of 160 nm. These restrictions are enforced during the optimization via a penalty term appended to the FOM.

The FOM for the optimization is given by expression \ref{eq:refname6} below where $\eta$ is given by equation \ref{eq:refname2}, $Q$ is the total number of wavelengths of interest, $\lambda_q$ denotes the specific wavelength of interest, and $p(\vec{x})$ is a penalty term as a function of the geometric parameters $\vec{x}$.

\begin{equation}
FOM=\sum_{q=1}^{Q}\left[\eta\left(\lambda_q\right)-p\left(\vec{x}\right)\right]
\label{eq:refname6}
\end{equation}

The penalty term is implemented through an approximation of the rectangular function given by:

\begin{equation}
p\left(\vec{x}\right)=A\left(\frac{1}{1+e^{-k\vec{x}}}+y_0\right)
\label{eq:refname7}
\end{equation}

\begin{figure*}[ht]
\centering
\includegraphics[width=\linewidth]{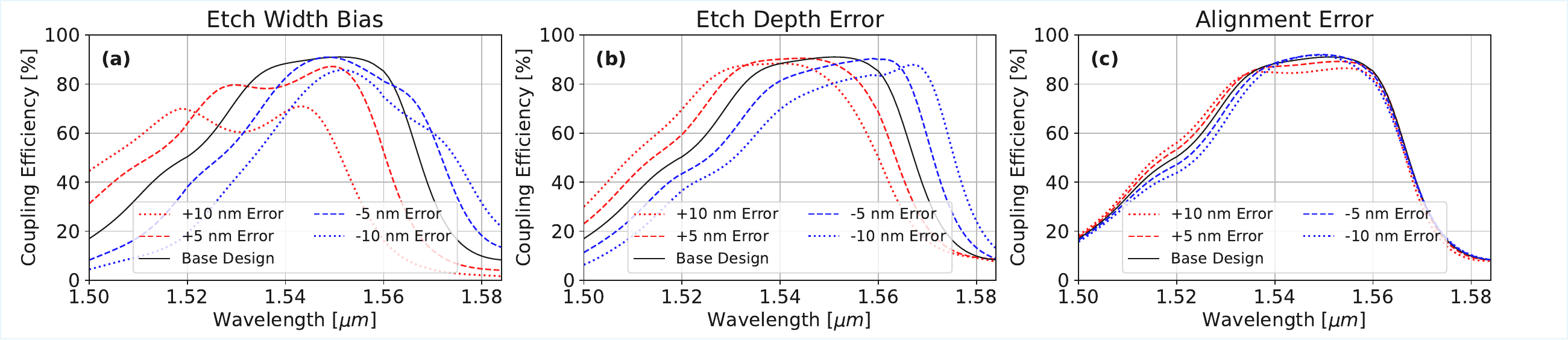}
\caption{Simulated impact of fabrication variations for \textbf{(a)} etch width bias, \textbf{(b)} etch depth error of the partially etched scattering sites, and \textbf{(c)} lateral alignment error between the partially and fully etched scattering sites. Etch width bias results in a reduction to the peak efficiency, a shift in the center wavelength, and for positive bias a broadening of the spectrum of a TWIG coupler. The dominant effect of an etch depth error is a shift in the central wavelength. These optimized devices are robust to alignment error and experience little affect over a wide range of errors.}
\label{fig:fabTolSim}
\end{figure*}

where $A$ is the scale of the penalty term, $k$ determines the steepness of the roll-off, and $y_0$ is the cut-off of the rectangular function which is set by the minimum feature size. The optimization is performed over three separate stages with increasing severity and steepness to the penalty term. This is done to enable a wide search of the parameter space before narrowing around a local optimum. Following the completion of all three stages, we predict a peak coupling efficiency of 91 $\%$ for the fundamental quasi-TE mode of the silicon waveguide. For this device, we simulate a 1 dB bandwidth of 34 nm spanning from 1530 nm to 1564 nm covering nearly the entire c-band. The performance and geometry of the optimized grating are given in figure \ref{fig:Simulations}. 

It can be seen that several of the local periods of the optimized design share similarities with designs discussed in \cite{Watanabe2017}, which investigates blazed gratings employing a nanopillar and L-shaped structure. It is the case that such a geometry is a subset of the TWIG coupler which requires that the partially etched sections overlap with the fully etched sections. Here, we solve and optimize a more generalized geometry, allowing for more degrees of freedom with which to satisfy the phase matching conditions for high directionality.

From this optimized design, we generate two variations of the TWIG coupler: a one-dimensional grating that only varies along the direction of propagation and a two-dimensional focusing grating coupler. The one-dimensional grating coupler maintains a constant width of 13 $\upmu${m} over the length of the grating and employs a 150 $\upmu${m} long adiabatic taper to transition between the width of the grating and a single mode waveguide. The two-dimensional grating coupler varies radially, introducing a curved phase profile which focuses the coupled light to a single mode waveguide at the origin of the radial etches \cite{Cheng2020}. The radial etches span an angle of $22^\circ$ and begin after an initial offset of 24.35 $\upmu${m} in order to achieve an approximate 10.4 $\upmu${m} MFD. The two-dimensional focusing coupler reduces the length required to transition from the grating mode to a single mode waveguide. The required footprint of the focusing TWIG is 41 $\upmu${m} $\times$ 16 $\upmu${m} while the one-dimensional grating requires a footprint of 150 $\upmu${m} $\times$ 13 $\upmu${m}. Figure \ref{fig:Simulations}\color{blue}d\color{black}\ and \ref{fig:Simulations}\color{blue}e\color{black}\ displays the two geometries with the simulated grating modes overlaid. 


\section{Simulated Fabrication Tolerance}

Variations in the fabrication process lead to inconsistencies between the designed device geometry and the final device. Here we investigate the impact of three different types of fabrication variation on the performance of our optimized grating couplers: etch width bias, etch depth error, and lateral alignment error between the lithography steps defining the partially etched scattering sites and fully etched scattering sites. For simplicity, etch width bias is treated as a constant dilation or shrinking for each scattering site in the grating region. Etch depth error is investigated by slightly adjusting the depth of the partial etch layer $H_3$, while it is assumed that a full etch of 220 nm can be reliably achieved. Alignment error is introduced by shifting all of the partially etched scattering sites together relative to the fully etched scattering sites. 

FDTD simulations of etch width bias reveal that errors on the order of $\pm$10 nm have a significant impact on the coupling spectrum of the device, both in terms of the peak coupling efficiency and bandwidth. We observe that for positive etch width bias (over-etch) the spectrum is broadened and shifted toward shorter wavelengths, while the peak efficiency is reduced. A negative bias (under-etch) narrows the spectrum while shifting it toward longer wavelengths and again reducing the peak efficiency. We observe an offset in the depth of the partially etched layer introduces a shift of the central wavelength while having only a small impact on the peak coupling efficiency and bandwidth. Increasing the etch depth shifts the spectrum toward shorter wavelengths, while reducing the etch depth shifts the spectrum toward longer wavelengths. We observe that alignment errors on the scale of $\pm$10 have very little impact on the device efficiency, bandwidth, or central wavelength. Figure \ref{fig:fabTolSim} above displays the expected impact of each type of fabrication variability on the optimized grating structure.


\section{Experimental Results}

\begin{figure}[ht!]
    \centering
    \begin{subfigure}[t]{0.5\linewidth}
        \centering
        \includegraphics[height=1.2in]{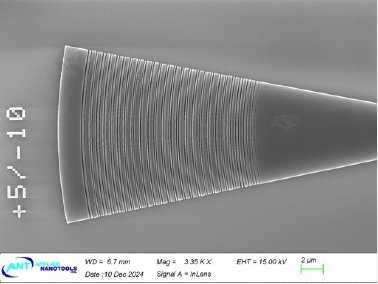}
    \end{subfigure}%
    ~ 
    \begin{subfigure}[t]{0.5\linewidth}
        \centering
        \includegraphics[height=1.2in]{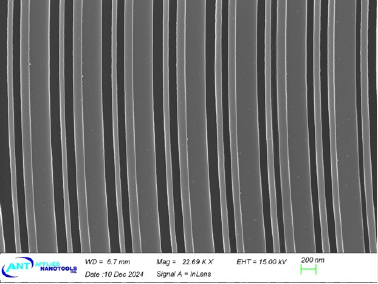}
    \end{subfigure}
    \caption{SEM images of the focusing TWIG coupler provided by Applied Nanotools.}
    \label{fig:SEM}
\end{figure}

The devices described above were submitted to the commercial foundry Applied Nanotools for fabrication. Applied Nanotools employs electron beam lithography to reliably achieve feature sizes down to 70 nm, ideal for low volume applications. As these designs require feature sizes below 100 nm, they are not compatible with the 193 nm process node typically used for integrated photonics multi-project wafer runs. However, increasingly advanced optical lithography is being pursued for integrated photonic applications, reducing the minimum feature sizes provided by commercial foundries. Figure \ref{fig:SEM} displays SEM micrographs of the fabricated couplers. 

Our test structures consist of an input TWIG coupler, a straight single mode waveguide, 500 nm $\times$ 220 nm, with a length of 1.03 cm, and an identical TWIG coupler rotated $180^\circ$ to act as an output coupler. We use an HP81680A tunable laser which operates in and around the c-band as our input source. The laser is fiber coupled to a polarization maintaining circulator, which enables us to monitor back reflections from the structure. The output of the circulator is coupled to the chip using a polarization maintaining fiber that has been cleaved and is held within a brass ferrule. The brass ferrule itself is held within a rotatable mount enabling us to tune the input polarization relative to the TWIG coupler. A cleaved SMF28 fiber is used at the output coupler before detection at an HP81531A photodiode. Alignment of the input and output fibers is managed by separate Thorlabs Nanomax 600 series 6 axis stages. Additionally, an index matching fluid (n = 1.444) has been used here to alleviate Fresnel reflections between the cladding/air and fiber/air interfaces. 

To account for possible fabrication errors, we have generated an array of TWIG couplers where each device in the array is designed to compensate for an alignment error, an etch width bias, or a combination of the two. An array of 15 devices is generated for each of the two variations accounting for etch width biases of scale 0 nm, $\pm$5 nm, $\pm$10 nm as well as alignment errors of scale 0 nm, $\pm$ 10 nm. Compensating for variations in etch depth would require re-optimization of the grating designs and as such we have elected not to include this process variation in our array.

To extract the efficiency of each TWIG coupler from an optical transmission measurement, we must first account for the propagation losses associated with the single mode waveguide. The experimental coupling efficiency is given by the function below where $P_{out}$ is the detected power, $P_{in}$ is the input power, and $T_{wg}$ is the transmission efficiency of the waveguide.

\begin{equation}
\eta_{GC}=\sqrt{\frac{P_{out}}{P_{in}T_{wg}}}
\label{eq:refname9}
\end{equation}

\begin{figure}[ht]
\centering
\includegraphics[width=\linewidth]{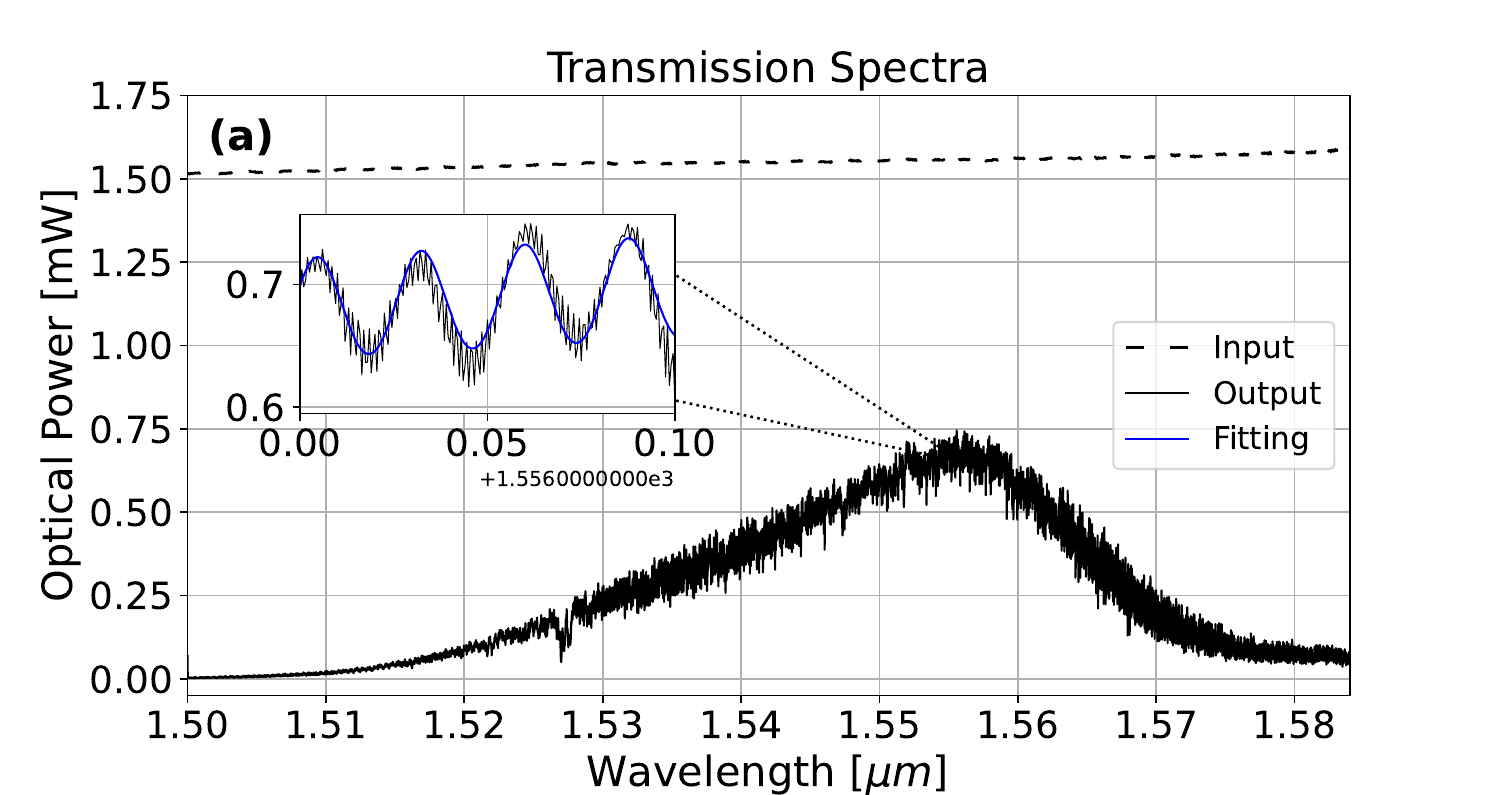}
\includegraphics[width=\linewidth]{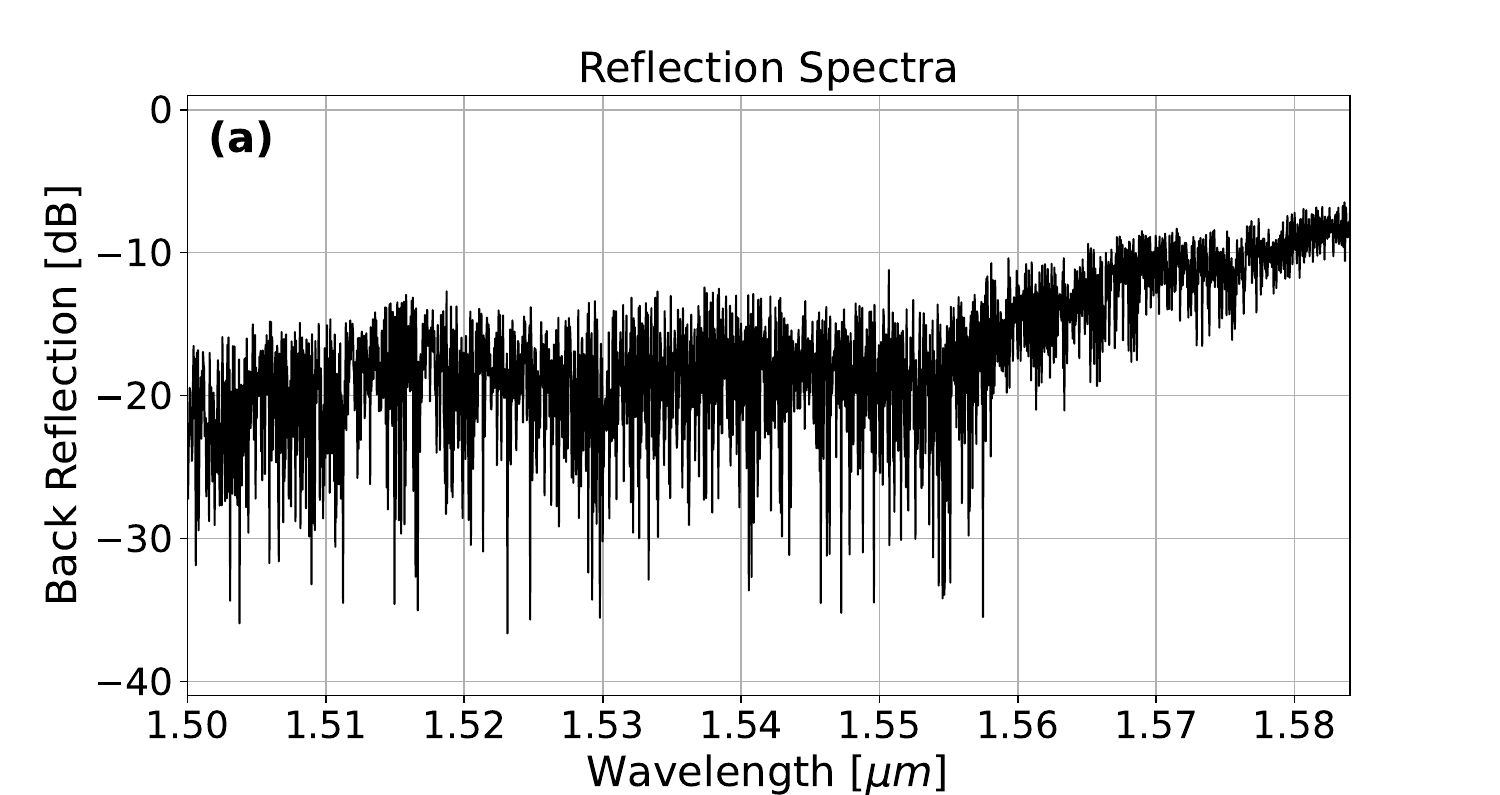}
\caption{\textbf{(a)} Transmission and \textbf{(b)} reflection spectra of the highest performing one-dimensional TWIG coupler. A waveguide propagation loss of -1.35 dB/cm is estimated from an analysis of the Fabry-Perot resonances.}
\label{fig:Transmission}
\end{figure}

We have estimated the propagation losses of our single mode waveguides from the Fabry-Perot resonances observed during the measurement of the highest performing one-dimensional TWIG coupler \cite{Hofstetter1997, Feuchter1994}. The Fabry-Perot resonances observed are a result of weak back-reflections at the waveguide/coupler interface. The measured transmission spectrum is given in figure \ref{fig:Transmission}\color{blue}\textbf{(a)}\color{black}\ where a maximum transmission of 47.8 $\%$ is observed at a wavelength of 1555.5 nm. The free spectral range of the Fabry-Perot resonance is 0.029 nm, which corresponds to a cavity length of 1.03 cm, the separation between the two gratings. 

The measured reflection spectrum is given in figure \ref{fig:Transmission}\color{blue}\textbf{(b)}\color{black}\ exhibiting a complimentary Fabry-Perot resonance with a similar free spectral range. At short wavelengths, we observe minimal back reflections on the order of -20 dB which increase sharply as the wavelength increases beyond 1560 nm. The measured back reflections around the peak coupling wavelength are observed to fluctuate between 0.1$\%$ and 4.5$\%$. This is of similar order of magnitude to back reflections measured for other techniques used to reduce reflections in perfectly vertical grating couplers \cite{Roelkens2007}. In this work back reflections were not considered in the FOM used for optimization. In future work, the FOM may be adjusted to explicitly minimize back reflections albeit with some trade off to the peak coupling efficiency.

To estimate the propagation losses, we perform a fitting of the measured transmission spectra to the analytical solution of a Fabry-Perot cavity with propagation losses. Within this fitting, the power reflectivity of the gratings is left as a fitting parameter. The optimal fitting parameters suggest a propagation loss of -1.35 dB/cm and a power reflectivity of 4.35$\%$, which are in good agreement with the average value reported by Applied Nanotools (-1.2 dB/cm) and the measured back reflections.



\begin{figure}[ht]
\centering
\includegraphics[width=\linewidth]{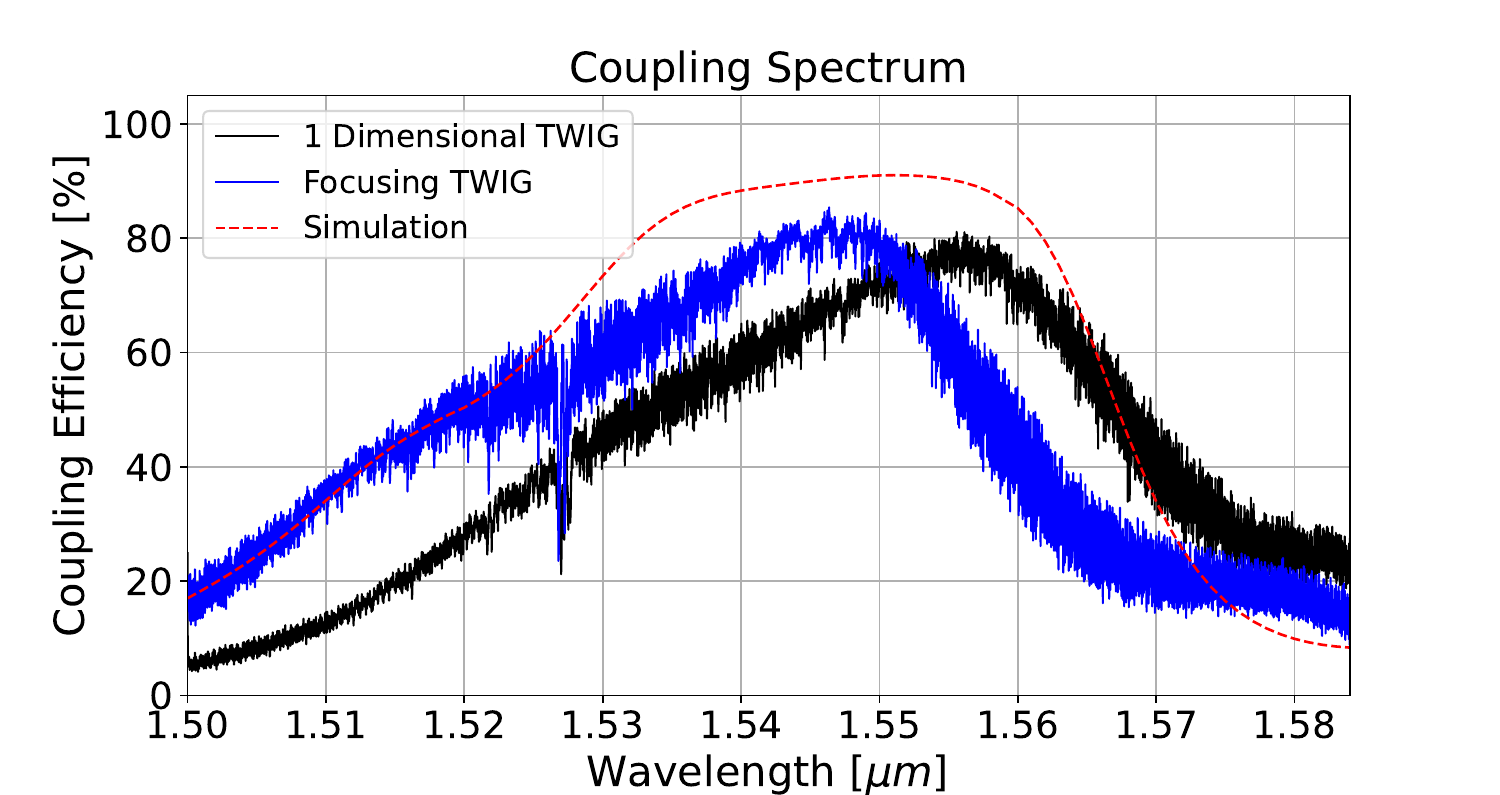}
\caption{Experimental coupling efficiency of the highest performing one dimensional and focusing TWIG couplers, each exhibiting sub-decibel insertion losses.}
\label{fig:CouplingEfficiency}
\end{figure}

According to equation \ref{eq:refname9}, from these results, we calculate the peak coupling efficiency of the one-dimensional TWIG coupler to be 81.1$\%$ (-0.91 dB) at a central wavelength of 1555.5 nm. The measured 1 dB bandwidth is 20 nm for this device. Similar measurements of the highest performing focusing TWIG coupler reveal a peak coupling efficiency of 85.4$\%$ (-0.69 dB) at a center wavelength of 1546.4 nm. We find a similar 1 dB bandwidth of 20 nm for the highest performing focusing TWIG coupler. The spectrum of each of these devices is plotted in figure \ref{fig:CouplingEfficiency} along with their simulated performance. We observe that there is a reduction of the peak efficiency and bandwidth of both devices compared to simulated values, however there is good agreement overall with the expected performance. 

Both devices exhibit an abrupt decrease in performance around 1528 nm. We believe this to be a measurement artifact associated with our testing station. This is indicated by the observation that each structure exhibits a similar drop at the same wavelength, 1528 nm, regardless of the shift to the peak coupling wavelength. 

The one-dimensional TWIG coupler plotted in figure \ref{fig:CouplingEfficiency} has been designed with a -5 nm etch width bias such that it compensates for a 5 nm over-etch. Additionally, this device has been designed for a -10 nm alignment error correction. The highest performing focusing TWIG coupler has been designed with a -10 nm etch width bias correction and a -10 nm alignment error. The difference in etch width bias between these two highest performing devices may be a result of local etch rate variations caused by the difference in geometry.

\begin{figure}[ht]
\centering
\includegraphics[width=\linewidth]{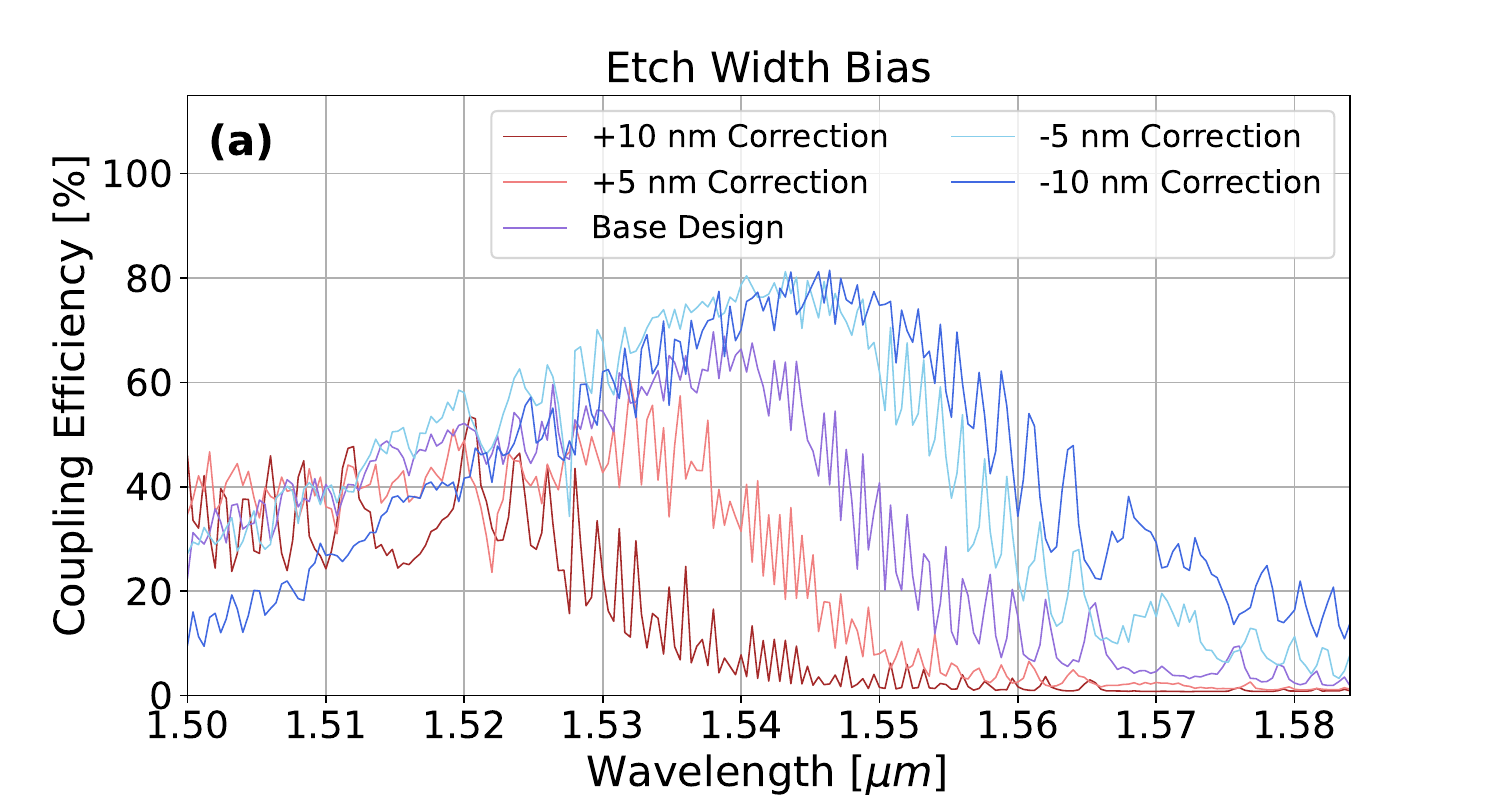}
\includegraphics[width=\linewidth]{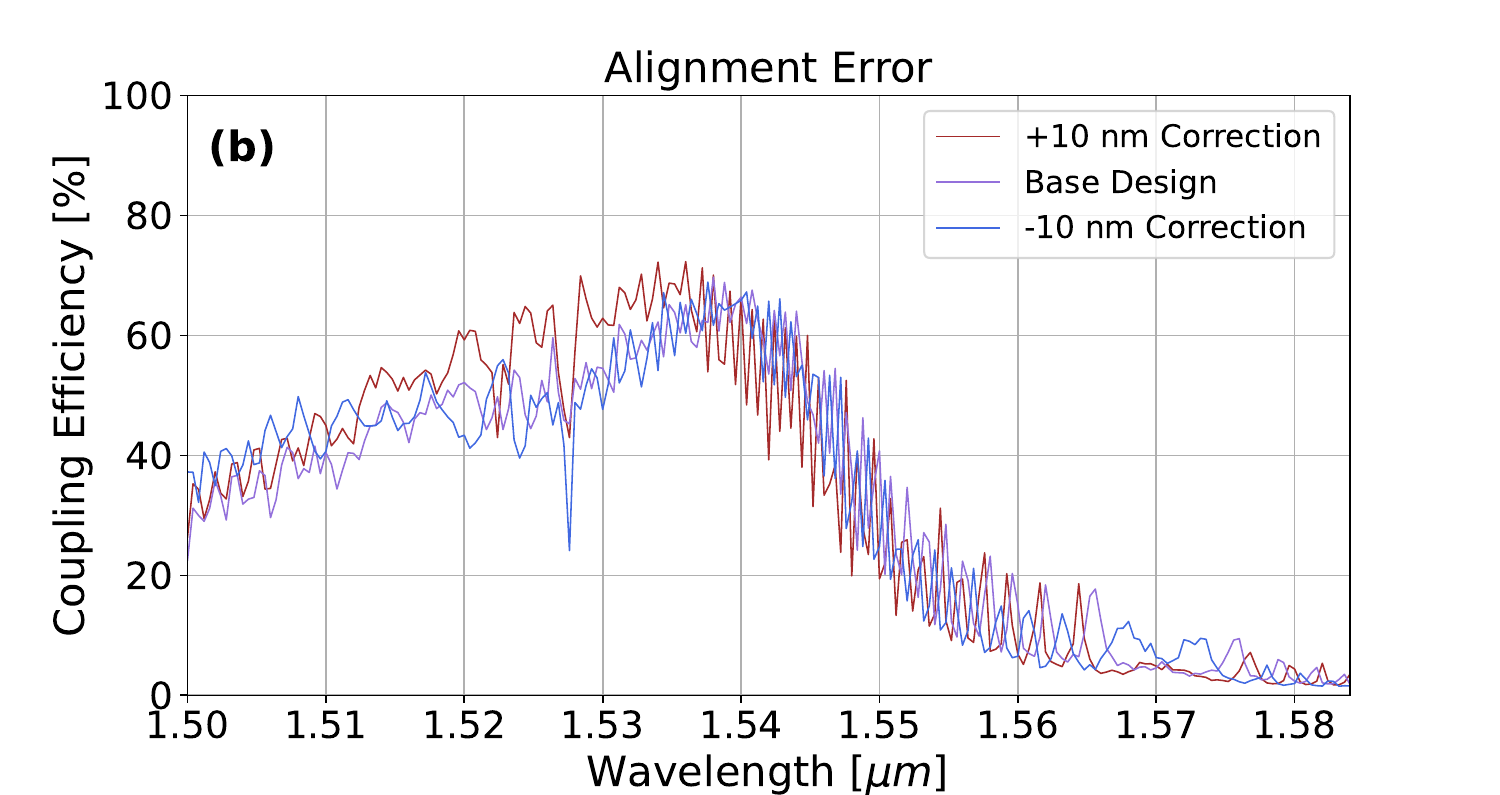}
\caption{Experiential coupling efficiency of focusing TWIG couplers demonstrating the impact of \textbf{(a)} etch width bias in the absence of alignment error and \textbf{(b)} alignment error in the absence of etch width bias.}
\label{fig:FabTolExp}
\end{figure}

Figure \ref{fig:FabTolExp} shows the measured coupling spectra of several of the remaining focusing TWIG couplers, displaying the impact of each type of fabrication variation. The wavelength resolution of the sweep here has been reduced for the convenience of viewing the closely overlapping measurements. We see good agreement between the experimental data and simulation with regard to both types of fabrication errors. We can see from panel \ref{fig:FabTolExp}\color{blue}a\color{black}\ that a positive etch width bias causes the spectra to broaden and shift toward shorter wavelengths while reducing peak coupling efficiency. We observe that a -10 nm correction exhibits the best performance among the devices with no overlap error. This indicates that the true etch width bias is close to -10 nm. From panel \ref{fig:FabTolExp}\color{blue}\textbf{b}\color{black}\ we can see that, regardless of alignment error, each of the devices performs similarly, demonstrating robustness to this type of fabrication variation. 

\begin{figure}[ht]
\centering
\includegraphics[width=\linewidth]{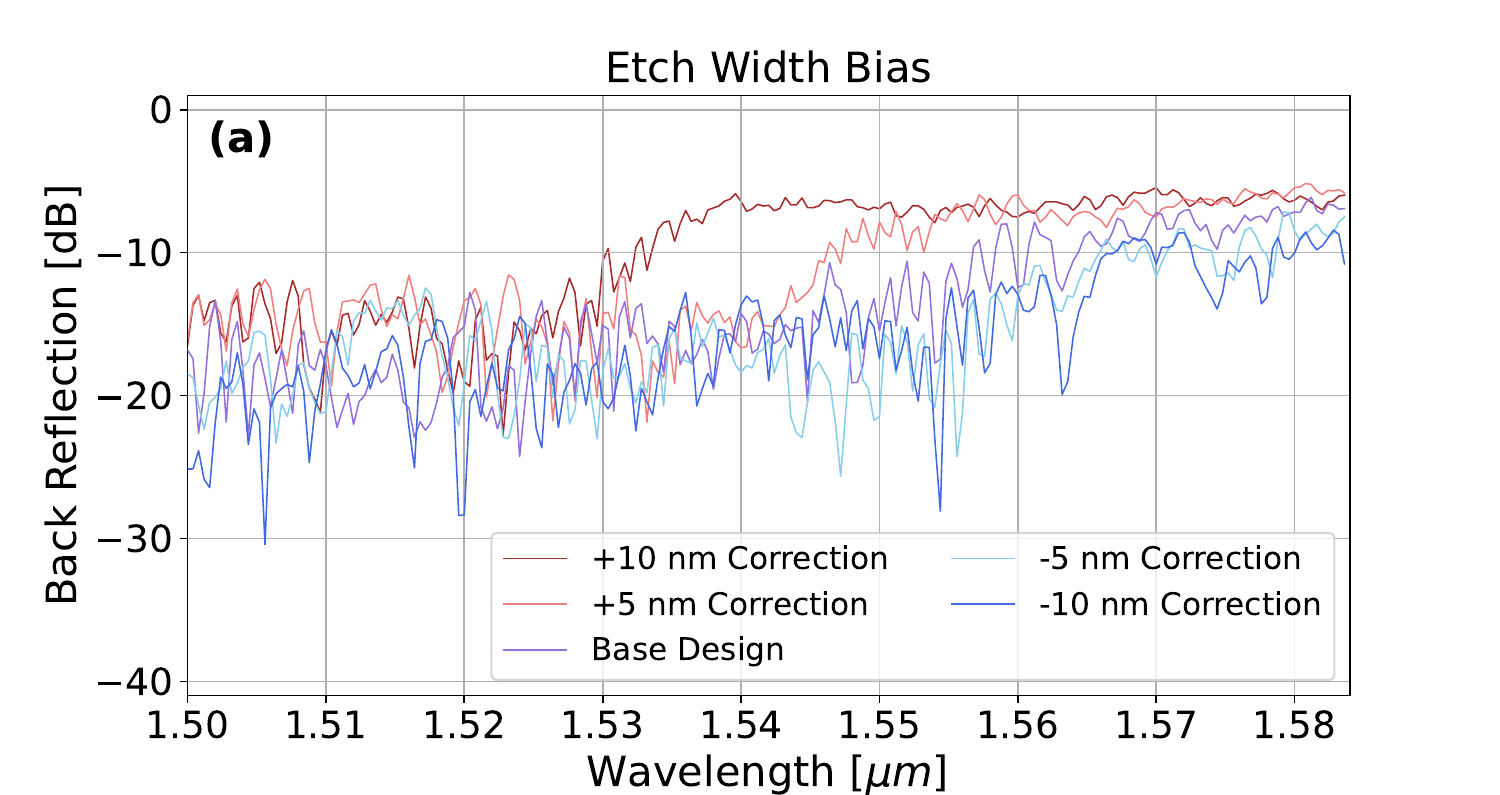}
\includegraphics[width=\linewidth]{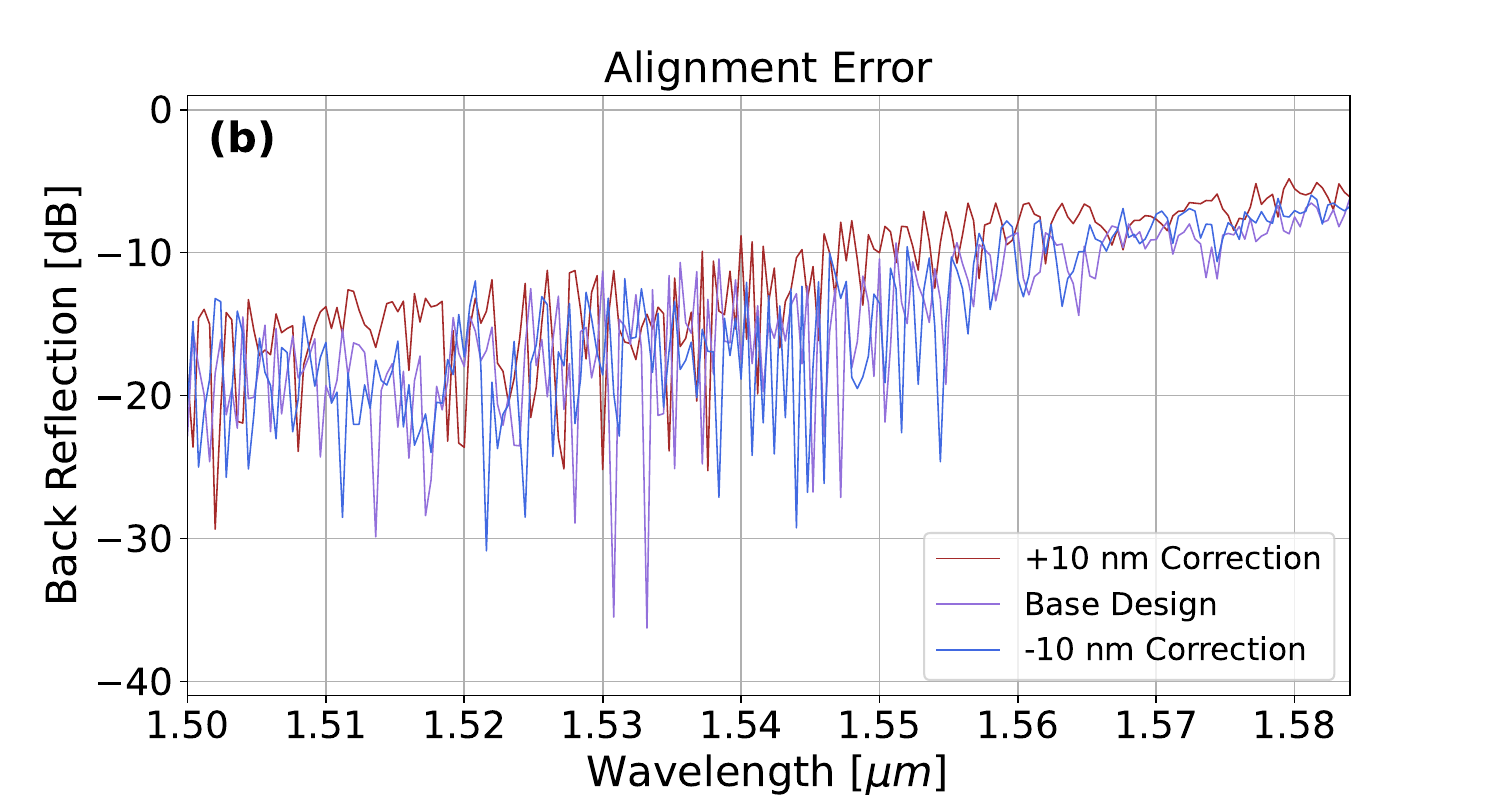}
\caption{Experiential reflection spectra of focusing TWIG couplers demonstrating the impact of \textbf{(a)} etch width bias in the absence of alignment error and \textbf{(b)} alignment error in the absence of etch width bias.}
\label{fig:FabTolExpB}
\end{figure}

Figure \ref{fig:FabTolExpB}\color{blue}\textbf{a}\color{black}\ provides the measured back reflection spectra that result from an etch width bias in the absence of alignment errors.
We observe an identical trend to that of the coupling efficiency. A positive etch width bias results in an increase in the measured back reflection intensity as well as a blue-shift of the center wavelength. Figure \ref{fig:FabTolExpB}\color{blue}\textbf{b}\color{black}\  displays the measured back reflection that results from alignment errors in the absence of etch width bias corrections. We similarly observe that the measured back reflection varies little for alignment errors on the order of $\pm$10 nm.

For potential etch width errors spanning $\pm$ 10 nm, we observe that the peak wavelength shifts by approximately 30 nm and the peak coupling efficiency is reduced from 85$\%$ to 50$\%$. For the same range of errors, the measured back reflections undergo a similar shift. For a +10 nm etch width bias, we observe back reflections as high as -20$\%$ (-7 dB) for wavelengths longer than 1540 nm. The observed phenomenon is likely caused by the gradual violation of the phase matching conditions required for high directionality \cite{Valdez2024}. These results suggest fairly stringent tolerances with regard to etch width bias; however, we observe that the device is robust to variations in the alignment between the full etch and partial etch steps. In future work, additional terms may be appended to the FOM to reduce the variability of device performance with respect to etch width bias.

\section{Conclusion}

We have designed, fabricated, via the commercial foundry Applied Nanotools, and experimentally validated the performance of high-efficiency grating couplers operating at normal incidence. By applying the principles of three-wave interaction gratings, we are able to optimize for the destructive interference of the typical loss mechanisms challenging perfectly vertical grating couplers: back reflections and substrate losses. Through careful design of the figure of merit we are able to balance the trade-offs between peak efficiency, bandwidth, and fabrication limitations. Our optimized designs predict a peak coupling efficiency of 91$\%$ with a 1 dB bandwidth of 34 nm spanning nearly the full C-band. From these designs we generate a one-dimensional TWIG coupler with a footprint of 150 $\upmu${m} $\times$ 13 $\upmu${m} and a focusing TWIG coupler with a reduced footprint of 41 $\upmu${m} $\times$ 16 $\upmu${m}. We measure peak coupling efficiencies of 81.1 $\%$ (-0.91 dB) and 85.4 $\%$ (-0.69 dB) for the one-dimensional and focusing devices respectively and 20 nm 1 dB bandwidth for both devices. To the best of our knowledge, these results represent the highest coupling efficiency for grating couplers designed in a standard 220 nm SOI platform, operating at normal incidence without the use of additional material layers such as bottom side reflectors or material overlays.

Adoption of these devices will benefit a wide range of applications of integrated photonics spanning the fields of optical communication, non-linear optics, spectroscopy, quantum optics, and astro-photonics. These devices offer a practical alternative to edge couplers in applications that are sensitive to insertion loss with the additional benefits associated with grating couplers: wider lateral alignment tolerance, convenient wafer-level testing, and flexible placement across the surface of a wafer. Operation at normal incidence enables coupling to vertically oriented components such as VCSEL and multicore fibers, both of which will play important roles in the scaling of optical interconnects. Extension of this method into new material platforms, wavelength ranges, and beam profiles provides a pathway to efficient coupling between any integrated photonic waveguides and fiber or free-space systems.

\begin{backmatter}
\bmsection{Funding} Ames Research Center (80NSSC24M0033); Stanford Engineering.

\bmsection{Acknowledgment} We thank the researchers Dan Sirbu, Kevin Fogarty, Ruslan Belikov, Rachel Morgan, and Eduardo Bendek-Selman from the Astrophysics branch of NASA AMES for many insightful discussions during the course of this work.



\bmsection{Disclosures} The authors declare no conflicts of interest.








\bmsection{Data availability} Data underlying the results presented in this paper are not publicly available at this time but may be obtained from the authors upon reasonable request.



\end{backmatter}




\bibliography{sample}

\begin{thebibliography}{10}
\newcommand{\enquote}[1]{``#1''}

\bibitem{Valdez23}
F.~Valdez, V.~Mere, X.~Wang, and S.~Mookherjea, \enquote{Integrated o- and c-band silicon-lithium niobate mach-zehnder modulators with 100 ghz bandwidth, low voltage, and low loss,} {\protect\JournalTitle{Opt. Express}} \textbf{31}, 5273--5289 (2023).

\bibitem{Shekhar2024}
S.~Shekhar, W.~Bogaerts, L.~Chrostowski, \emph{et~al.}, \enquote{Roadmapping the next generation of silicon photonics,} {\protect\JournalTitle{Nature Communications}} \textbf{15}, 751 (2024).

\bibitem{Li2022}
A.~Li, C.~Yao, J.~Xia, \emph{et~al.}, \enquote{Advances in cost-effective integrated spectrometers,} {\protect\JournalTitle{Light: Science \& Applications}} \textbf{11}, 174 (2022).

\bibitem{miller2025}
D.~A.~B. Miller, C.~Roques-Carmes, C.~G. Valdez, \emph{et~al.}, \enquote{Universal programmable and self-configuring optical filter,}  (2025).

\bibitem{Bates2021}
R.~Baets, \enquote{Silicon-photonics-based spectroscopic sensing for environmental monitoring and health care,} in \emph{2021 Optical Fiber Communications Conference and Exhibition (OFC),}  (2021), pp. 1--42.

\bibitem{Psiquantum}
K.~Alexander, A.~Benyamini, D.~Black, \emph{et~al.}, \enquote{A manufacturable platform for photonic quantum computing,} {\protect\JournalTitle{Nature}} \textbf{641}, 876--883 (2025).

\bibitem{Luo2023}
W.~Luo, L.~Cao, Y.~Shi, \emph{et~al.}, \enquote{Recent progress in quantum photonic chips for quantum communication and internet,} {\protect\JournalTitle{Light: Science \& Applications}} \textbf{12}, 175 (2023).

\bibitem{Sirbu2024}
D.~Sirbu, R.~Belikov, K.~Fogarty, \emph{et~al.}, \enquote{{AstroPIC: near-infrared photonic integrated circuit coronagraph architecture for the Habitable Worlds Observatory},} in \emph{Space Telescopes and Instrumentation 2024: Optical, Infrared, and Millimeter Wave,}  vol. 13092 L.~E. Coyle, S.~Matsuura, and M.~D. Perrin, eds., International Society for Optics and Photonics (SPIE, 2024), p. 130921T.

\bibitem{Morgan2025}
R.~{Morgan}, C.~{Valdez}, A.~{Kroo}, \emph{et~al.}, \enquote{{Status and Predicted Performance for the AstroPIC Integrated Photonic Coronagraph},} in \emph{American Astronomical Society Meeting Abstracts,}  vol. 245 of \emph{American Astronomical Society Meeting Abstracts} (2025), p. 234.07.

\bibitem{Glint}
M.-A. Martinod, B.~Norris, P.~Tuthill, \emph{et~al.}, \enquote{Scalable photonic-based nulling interferometry with the dispersed multi-baseline glint instrument,} {\protect\JournalTitle{Nature Communications}} \textbf{12}, 2465 (2021).

\bibitem{Marchetti19}
R.~Marchetti, C.~Lacava, L.~Carroll, \emph{et~al.}, \enquote{Coupling strategies for silicon photonics integrated chips \[invited\],} {\protect\JournalTitle{Photon. Res.}} \textbf{7}, 201--239 (2019).

\bibitem{Mu2020}
X.~Mu, S.~Wu, L.~Cheng, and H.~Fu, \enquote{Edge couplers in silicon photonic integrated circuits: A review,} {\protect\JournalTitle{Applied Sciences}} \textbf{10} (2020).

\bibitem{Yuan2025}
Q.~Yuan, A.~Peczek, J.~Frankel, \emph{et~al.}, \enquote{Fully automated wafer-level edge coupling measurement system for silicon photonics integrated circuits,} {\protect\JournalTitle{IEEE Transactions on Semiconductor Manufacturing}} \textbf{38}, 168--177 (2025).

\bibitem{Barwicz2019}
T.~Barwicz, B.~Peng, R.~Leidy, \emph{et~al.}, \enquote{Integrated metamaterial interfaces for self-aligned fiber-to-chip coupling in volume manufacturing,} {\protect\JournalTitle{IEEE Journal of Selected Topics in Quantum Electronics}} \textbf{25}, 1--13 (2019).

\bibitem{Wang2016}
J.~Wang, Y.~Xuan, C.~Lee, \emph{et~al.}, \enquote{Low-loss and misalignment-tolerant fiber-to-chip edge coupler based on double-tip inverse tapers,} in \emph{2016 Optical Fiber Communications Conference and Exhibition (OFC),}  (2016), pp. 1--3.

\bibitem{Cheng2020}
L.~Cheng, S.~Mao, Z.~Li, \emph{et~al.}, \enquote{Grating couplers on silicon photonics: Design principles, emerging trends and practical issues,} {\protect\JournalTitle{Micromachines}} \textbf{11} (2020).

\bibitem{Hooten2020}
S.~Hooten, T.~V. Vaerenbergh, P.~Sun, \emph{et~al.}, \enquote{Adjoint optimization of efficient cmos-compatible si-sin vertical grating couplers for dwdm applications,} {\protect\JournalTitle{Journal of Lightwave Technology}} \textbf{38}, 3422--3430 (2020).

\bibitem{Watanabe2017}
T.~Watanabe, M.~Ayata, U.~Koch, \emph{et~al.}, \enquote{Perpendicular grating coupler based on a blazed antiback-reflection structure,} {\protect\JournalTitle{Journal of Lightwave Technology}} \textbf{35}, 4663--4669 (2017).

\bibitem{Tong2018}
Y.~Tong, W.~Zhou, and H.~K. Tsang, \enquote{Efficient perfectly vertical grating coupler for multi-core fibers fabricated with 193\&\#x2009;\&\#x2009;nm duv lithography,} {\protect\JournalTitle{Opt. Lett.}} \textbf{43}, 5709--5712 (2018).

\bibitem{Zhao2020}
Z.~Zhao and S.~Fan, \enquote{Design principles of apodized grating couplers,} {\protect\JournalTitle{J. Lightwave Technol.}} \textbf{38}, 4435--4446 (2020).

\bibitem{Bozzola2015}
A.~{Bozzola}, L.~{Carroll}, D.~{Gerace}, \emph{et~al.}, \enquote{{Optimising apodized grating couplers in a pure SOI platform to -05 dB coupling efficiency},} {\protect\JournalTitle{Optics Express}} \textbf{23}, 16289 (2015).

\bibitem{Marchetti2017}
R.~Marchetti, C.~Lacava, A.~Khokhar, \emph{et~al.}, \enquote{High-efficiency grating-couplers: Demonstration of a new design strategy,} {\protect\JournalTitle{Scientific Reports}} \textbf{7} (2017).

\bibitem{Zhang2019}
Z.~Zhang, X.~Chen, Q.~Cheng, \emph{et~al.}, \enquote{High-efficiency apodized bidirectional grating coupler for perfectly vertical coupling,} {\protect\JournalTitle{Opt. Lett.}} \textbf{44}, 5081--5084 (2019).

\bibitem{Laere2007}
F.~V. Laere, G.~Roelkens, M.~Ayre, \emph{et~al.}, \enquote{Compact and highly efficient grating couplers between optical fiber and nanophotonic waveguides,} {\protect\JournalTitle{J. Lightwave Technol.}} \textbf{25}, 151--156 (2007).

\bibitem{Vermeulen2010}
D.~Vermeulen, S.~Selvaraja, P.~Verheyen, \emph{et~al.}, \enquote{High-efficiency fiber-to-chip grating couplers realized using an advanced cmos-compatible silicon-on-insulator platform,} {\protect\JournalTitle{Opt. Express}} \textbf{18}, 18278--18283 (2010).

\bibitem{Hooten2022}
S.~Hooten, M.~Jain, T.~Van~Vaerenbergh, \emph{et~al.}, \enquote{Inverse-designed dual layer c-si/sin vertical grating couplers tested on 300mm wafers,} in \emph{2022 27th OptoElectronics and Communications Conference (OECC) and 2022 International Conference on Photonics in Switching and Computing (PSC),}  (2022), pp. 1--4.

\bibitem{Benedikovic2015}
D.~Benedikovic, C.~Alonso-Ramos, P.~Cheben, \emph{et~al.}, \enquote{High-directionality fiber-chip grating coupler with interleaved trenches and subwavelength index-matching structure,} {\protect\JournalTitle{Opt. Lett.}} \textbf{40}, 4190--4193 (2015).

\bibitem{Chen2017}
X.~Chen, D.~J. Thomson, L.~Crudginton, \emph{et~al.}, \enquote{Dual-etch apodised grating couplers for efficient fibre-chip coupling near 1310 nm wavelength,} {\protect\JournalTitle{Opt. Express}} \textbf{25}, 17864--17871 (2017).

\bibitem{Dezfouli2020}
M.~K. Dezfouli, Y.~Grinberg, D.~Melati, \emph{et~al.}, \enquote{Perfectly vertical surface grating couplers using subwavelength engineering for increased feature sizes,} {\protect\JournalTitle{Opt. Lett.}} \textbf{45}, 3701--3704 (2020).

\bibitem{Tian2022}
Z.-T. Tian, Z.-P. Zhuang, Z.-B. Fan, \emph{et~al.}, \enquote{High-efficiency grating couplers for pixel-level flat-top beam generation,} {\protect\JournalTitle{Photonics}} \textbf{9} (2022).

\bibitem{Zhou2022}
X.~Zhou and H.~K. Tsang, \enquote{Optimized shift-pattern overlay for high coupling efficiency waveguide grating couplers,} {\protect\JournalTitle{Opt. Lett.}} \textbf{47}, 3968--3971 (2022).

\bibitem{Zhou2023}
X.~Zhou and H.~K. Tsang, \enquote{Photolithography fabricated sub-decibel high-efficiency silicon waveguide grating coupler,} {\protect\JournalTitle{IEEE Photonics Technology Letters}} \textbf{35}, 43--46 (2023).

\bibitem{Naoki2025}
N.~Tahara, S.~Nawa, R.~Taira, \emph{et~al.}, \enquote{{High efficiency silicon photonics grating coupler compatible to standard foundry service},} in \emph{Silicon Photonics XX,}  vol. 13371 G.~T. Reed and J.~Bradley, eds., International Society for Optics and Photonics (SPIE, 2025), p. 133710C.

\bibitem{Michaels2018}
A.~Michaels and E.~Yablonovitch, \enquote{Inverse design of near unity efficiency perfectly vertical grating couplers,} {\protect\JournalTitle{Opt. Express}} \textbf{26}, 4766--4779 (2018).

\bibitem{Notaros2016}
J.~Notaros, F.~Pavanello, M.~T. Wade, \emph{et~al.}, \enquote{Ultra-efficient cmos fiber-to-chip grating couplers,} in \emph{2016 Optical Fiber Communications Conference and Exhibition (OFC),}  (2016), pp. 1--3.

\bibitem{Dai2015}
M.~Dai, L.~Ma, Y.~Xu, \emph{et~al.}, \enquote{Highly efficient and perfectly vertical chip-to-fiber dual-layer grating coupler,} {\protect\JournalTitle{Opt. Express}} \textbf{23}, 1691--1698 (2015).

\bibitem{Valdez2024}
C.~G. Valdez, S.~Pai, P.~Broaddus, and O.~Solgaard, \enquote{High-efficiency vertically emitting coupler facilitated by three wave interaction gratings,} {\protect\JournalTitle{Opt. Lett.}} \textbf{49}, 2373--2376 (2024).

\bibitem{Michaels2020}
A.~Michaels, M.~C. Wu, and E.~Yablonovitch, \enquote{Hierarchical design and optimization of silicon photonics,} {\protect\JournalTitle{IEEE Journal of Selected Topics in Quantum Electronics}} \textbf{26}, 1--12 (2020).

\bibitem{Hofstetter1997}
D.~Hofstetter and R.~L. Thornton, \enquote{Theory of loss measurements of fabry--perot resonators by fourier analysis of the transmission spectra,} {\protect\JournalTitle{Opt. Lett.}} \textbf{22}, 1831--1833 (1997).

\bibitem{Feuchter1994}
T.~Feuchter and C.~Thirstrup, \enquote{High precision planar waveguide propagation loss measurement technique using a fabry-perot cavity,} {\protect\JournalTitle{IEEE Photonics Technology Letters}} \textbf{6}, 1244--1247 (1994).

\bibitem{Roelkens2007}
G.~Roelkens, D.~V. Thourhout, and R.~Baets, \enquote{High efficiency grating coupler between silicon-on-insulator waveguides and perfectly vertical optical fibers,} {\protect\JournalTitle{Opt. Lett.}} \textbf{32}, 1495--1497 (2007).

\end{thebibliography}



\ifthenelse{\equal{\journalref}{aop}}{%
\section*{Author Biographies}
\begingroup
\setlength\intextsep{0pt}
\begin{minipage}[t][6.3cm][t]{1.0\textwidth} 
  \begin{wrapfigure}{L}{0.25\textwidth}
    \includegraphics[width=0.25\textwidth]{john_smith.eps}
  \end{wrapfigure}
  \noindent
  {\bfseries John Smith} received his BSc (Mathematics) in 2000 from The University of Maryland. His research interests include lasers and optics.
\end{minipage}
\begin{minipage}{1.0\textwidth}
  \begin{wrapfigure}{L}{0.25\textwidth}
    \includegraphics[width=0.25\textwidth]{alice_smith.eps}
  \end{wrapfigure}
  \noindent
  {\bfseries Alice Smith} also received her BSc (Mathematics) in 2000 from The University of Maryland. Her research interests also include lasers and optics.
\end{minipage}
\endgroup
}{}

\end{document}